\documentclass[lettersize,journal]{IEEEtran}
\usepackage{amsmath,amsfonts}
\usepackage{subcaption}
\usepackage{algorithm}
\usepackage{algpseudocode}
\usepackage{array}
\usepackage[caption=false,font=normalsize,labelfont=sf,textfont=sf]{subfig}
\usepackage{textcomp}
\usepackage[table,xcdraw]{xcolor}
\usepackage{stfloats}
\usepackage{url}
\usepackage{verbatim}
\usepackage{graphicx}
\usepackage{cite}
\usepackage{hhline}
\usepackage{nicematrix}
\usepackage{mathrsfs}
\usepackage{amsmath}
\usepackage{amssymb}
\usepackage{amsfonts}
\usepackage{array}
\usepackage{multirow}
\usepackage{tabularx}

\usepackage{epstopdf}
\usepackage{booktabs}

\usepackage{mathtools}

\begin{document}

\title{CUBA: Controlled Untargeted Backdoor Attack against Deep Neural Networks}

\author{Yinghao~Wu and Liyan~Zhang
\thanks{Yinghao Wu and Liyan Zhang are with the College of Computer Science and Technology, Nanjing University of Aeronautics and Astronautics, Nanjing, 211106, China (e-mail: wyh@nuaa.edu.cn; zhangliyan@nuaa.edu.cn).}
}

\markboth{Journal of \LaTeX\ Class Files,~Vol.~14, No.~8, August~2021}%
{Shell \MakeLowercase{\textit{et al.}}: A Sample Article Using IEEEtran.cls for IEEE Journals}


\maketitle

\begin{abstract}
Backdoor attacks have emerged as a critical security threat against deep neural networks in recent years.
The majority of existing backdoor attacks focus on targeted backdoor attacks, where trigger is strongly associated to specific malicious behavior.
Various backdoor detection methods depend on this inherent property and shows effective results in identifying and mitigating such targeted attacks.
However, a purely untargeted attack in backdoor scenarios is, in some sense, self-weakening, since the target nature is what makes backdoor attacks so powerful.
In light of this, we introduce a novel \underline{C}onstrained \underline{U}ntargeted \underline{B}ackdoor \underline{A}ttack (CUBA), which combines the flexibility of untargeted attacks with the intentionality of targeted attacks.
The compromised model, when presented with backdoor images, will classify them into random classes within a constrained range of target classes selected by the attacker.
This combination of randomness and determinedness enables the proposed untargeted backdoor attack to natively circumvent existing backdoor defense methods.
To implement the untargeted backdoor attack under controlled flexibility, we propose to apply logit normalization on cross-entropy loss with flipped one-hot labels.
By constraining the logit during training, the compromised model will show a uniform distribution across selected target classes, resulting in controlled untargeted attack.
Extensive experiments demonstrate the effectiveness of the proposed CUBA on different datasets.
\end{abstract}


\section{Introduction}\label{sec:intro}
\IEEEPARstart{D}{eep} Neural Networks (DNNs) show superior performance than traditional machine learning algorithms and thus have been widely used in various applications.
Despite their remarkable performance, the efficacy of deep learning models is heavily dependent on two critical factors: abundant hight quality training data and substantial computational resources.
This dependency imposes significant constraints on both research endeavors and practical implementations, particularly for individual researchers and small organizations with limited resources.
The computational expenses associated with training state-of-the-art deep learning models can be prohibitively high, often requiring specialized knowledge and extensive time investment.
Consequently, a growing trend has emerged wherein researchers, developers, and organizations opt to either outsource the training process to third-party platforms or directly deploy pre-trained models available from online repositories or other external sources.
While this approach alleviates the immediate burden of model training, it introduces new challenges and potential vulnerabilities to the deep learning ecosystem.
Users deploying such models often lack visibility into the training process, data sources, and potential manipulations that may have occurred during model development.
The opaque nature of deep learning models and lack of transparency on them create an environment conducive to various security threats, one of which is backdoor attack as a particularly insidious risk.

\begin{figure}[!t]
    \centering
    \includegraphics[width=\linewidth]{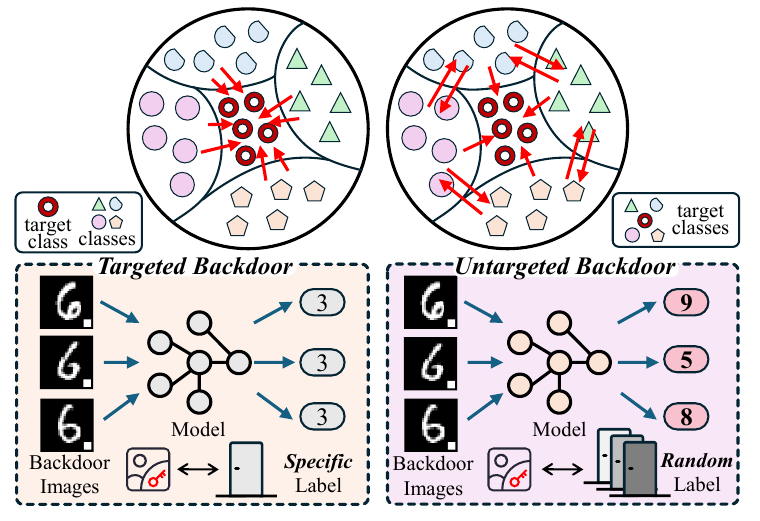}
    \caption{Difference between targeted and untargeted backdoor attacks. Red line indicates the image is misclassified from its ground-truth class to target class.}
    \label{fig:front}
\end{figure}

Backdoor attacks, wherein malicious attackers embed hidden functionalities into the model during the training phase, can remain undetected as the model performs well on standard tasks.
Since the inception of backdoor attacks with BadNets \cite{abs-1708-06733}, they have been inherently targeted in nature.
Through backdoor training, attackers specify target classes and force models to establish strong correlations between triggers and target classes while maintaining the model's normal classification performance.
This duality of stealthiness and deterministic behavior has made backdoor attacks particularly powerful.
Yet this very characteristic leaves distinctive traces in the model, especially in the feature space \cite{ZhongQZ22, TanS2020_BypassingBackdoor, ChenYZBS22}, providing defenders with leverage points for backdoor detection and removal.
This observation led us to explore the possibility of untargeted backdoor attacks, similar to adversarial examples \cite{GoodfellowSS14, Moosavi-Dezfooli17}, which is illustrated in Fig. \ref{fig:front}.
Nevertheless, a completely untargeted approach would sacrifice the strategic advantage of backdoor attacks' deterministic nature.
To this end, we explore a kind of controlled untargeted attack that combines both determinism and randomness, where backdoor inputs are expected to be misclassified randomly within a constrained set of classes.

\begin{table*}[htbp]
\caption{Comparison of Different Attack Paradigms}
\label{tab:attack_comparison}
\centering
\begin{tabularx}{\textwidth}{>{\centering\arraybackslash}p{3.5cm}c>{\centering\arraybackslash}X>{\centering\arraybackslash}X>{\centering\arraybackslash}X}
\toprule
Characteristics & Targeted Backdoor & Controlled Untargeted Backdoor & Universal Adversarial Perturbation \\
\midrule
Attack Stage & Training & Training & Inference \\
Model Modification & Yes & Yes & No \\
Trigger/Perturbation & Controllable & Controllable & Fixed but uncontrollable \\
Attack Scope & Single target & Controlled random & Uncontrolled random \\
Stealthiness & High & High & Limited \\
Persistence & Permanent & Permanent & Temporary \\
Success Rate & Higher & High & Moderate \\
Control Granularity & Fine-grained & Medium-grained & Coarse-grained \\
\bottomrule
\end{tabularx}
\end{table*}

In this paper, we propose a novel backdoor attack without a specific targeted class but still under certain control, called Controlled Untargeted Backdoor Attack (CUBA).
Different from traditional targeted backdoor attacks, CUBA does not establish a direct connection between the trigger and a specific target class.
Instead, when presented with backdoor images, the compromised model will randomly misclassify inputs into incorrect classes, but within a predefined subset of classes chosen by the attacker.
Specifically, there are two variants of our proposed CUBA: \textbf{Full Range Attack (FRA)} and \textbf{Narrow Range Attack (NRA)}.
For the former, the compromised model randomly misclassifies backdoor inputs into any incorrect class within the entire output space, excluding only the ground-truth class of the input.
For the latter, this variant constrains the misclassification to a fixed number of classes in the vicinity of the input's ground-truth class, excluding the true class itself.
The vicinity is defined based on a predetermined metric or class relationship.
This NRA allows one trigger to activate a set of random target classes, while still maintaining controlled characteristics.
However, ensuring a uniform probability distribution of misclassifications across the designated subset of classes, while excluding the true class, poses considerable difficulties.
Through extensive experimentation, we observe backdoor models tend to show overconfidence on backdoor images, reflecting the strong connection built between the backdoor and trigger.
This observation provides a crucial insight: by mitigating this overconfidence pattern, we can redistribute the model's confidence more evenly across selected classes.
Based on this intuition, we propose a two-pronged approach combining logit normalization and one-hot label flipping to regulate the confidence distribution prior to the cross-entropy loss computation. 
While the standard cross-entropy loss function is designed to minimize the divergence between predicted and ground-truth label distributions, our proposed label flipping mechanism deliberately reverses this optimization direction. 
Through this carefully designed training strategy, the model learns to generate predictions for backdoor images that maximize entropy with respect to the ground-truth class, effectively achieving controlled misclassification while preserving performance on clean inputs.
The main contributions of this paper are summarized as follows:
\begin{itemize}
  \item We propose a new \textit{controlled untargted backdoor attack} paradigm, which, to the best of authors' knowledge, is the first untargeted backdoor attack with controllability.
  This paradigm break the assumption of one-to-one mapping between backdoor and trigger, wherein one trigger can randomly activate multiple backdoor classes but within a predefined range.
  By distributing misclassifications across multiple classes, the attack becomes less conspicuous and harder to detect through statistical analysis.
  \item We develop an effective training mechanism that combines logit normalization with flipped one-hot encoding to achieve controlled randomization in backdoor predictions. This technique regulates the model's confidence distribution during training, ensuring uniform misclassification across selected classes while maintaining model performance on clean inputs. Our approach demonstrates that careful manipulation of the training objective can achieve sophisticated attack patterns that challenge existing defense assumptions.
\end{itemize}

The remainder of this paper is organized as follows:
Section \ref{sec:compare} compares the proposed backdoor attack methods with other attack paradigms.
Section \ref{sec:related_works} provides an overview of backdoor attack and defense methods. 
Section \ref{sec:proposed_method} presents the proposed controlled untargeted backdoor attack. 
Section \ref{sec:experiments} discusses the experimental setup and the results of the proposed methods. 
Finally, Section \ref{sec:conclusion} offers a comprehensive discussion, including limitations and potential improvements. 

\section{Comparison of Attack Paradigms}\label{sec:compare}
In this section, we present a systematic comparison between untargeted backdoor attacks, targeted backdoor attacks, and Universal Adversarial Perturbations (UAP). Table \ref{tab:attack_comparison} summarizes the key characteristics of these attack paradigms.

The fundamental distinction between backdoor attacks and adversarial attacks lies in their operational mechanism. Backdoor attacks manipulate the model during the training phase, permanently altering the model's parameters, while adversarial attacks operate during inference without modifying the model architecture or parameters. This distinction leads to several important implications for attack effectiveness and detectability.
While UAPs share some similarities with untargeted backdoor attacks, particularly in their ability to cause misclassification across multiple inputs using a fixed perturbation pattern, several key advantages make untargeted backdoor attacks more sophisticated and practical:
\begin{itemize}
  \item \textbf{Trigger Stealthiness:} Untargeted backdoor attacks can leverage advanced trigger generation techniques, including generative models, to create imperceptible or naturalistic triggers. In contrast, UAPs typically produce visible perturbations that are more easily detectable by human observers or automated detection systems.
  \item \textbf{Attack Controllability:} Our proposed controlled untargeted backdoor attack introduces a novel capability to constrain the misclassification space while maintaining randomness within specified class boundaries. This offers a unique balance between chaos and control that is unattainable with UAPs, where the misclassification pattern is inherently unpredictable.
  \item \textbf{Attack Persistence:} Backdoor attacks embed the vulnerability directly into the model's parameters, ensuring consistent behavior across different deployment scenarios. UAPs, being inference-time attacks, may require repeated application and are more susceptible to defensive preprocessing techniques.
  \item \textbf{Implementation Flexibility:} Backdoor triggers can be designed with various forms (e.g., physical patterns, digital watermarks, or semantic features), while UAPs are typically limited to additive pixel-level perturbations.
\end{itemize}

The controlled untargeted backdoor attack represents a significant advancement in adversarial machine learning, combining the benefits of traditional backdoor attacks with enhanced controllability and stealthiness. This paradigm enables attackers to achieve a precise balance between randomness and control, making it particularly challenging for existing defense mechanisms while maintaining practical applicability in real-world scenarios.
Furthermore, the ability to specify the scope of potential misclassifications while ensuring randomness within that scope represents a novel attack capability that neither targeted backdoors nor UAPs can achieve. This characteristic is particularly relevant in scenarios where complete predictability (as in targeted attacks) might be detectable, but entirely random misclassification (as in UAPs) might be too chaotic for the attacker's purposes.

\section{Related Work and Preliminaries}\label{sec:related_works}
\subsection{Related Work}
We believe all existing backdoor attacks are targeted attack \cite{Nguyen2021_WanetImperceptibleWarpingBased, SahaSP2020_HiddenTriggerBackdoor, TanS2020_BypassingBackdoor, Cheng2021_DeepFeatureSpace, Doan2021_LIRA, BlindBackdoorsDeep2021Bagdasaryan, BackdoorAttackImperceptible2021Doan, Li2021_samplespecific, marksman_DoanL022, inputaware_NguyenT20, dynamic_SalemWBMZ22, PoisonFrogsTargeted2018Shafahi, BppAttackStealthyEfficient2022Wang, zhu2019transferable, li2025feat, WangMGAZKAZA24, XingXBY24}.
Although there are attacks that claim they are dynamic attack \cite{dynamic_SalemWBMZ22}, feature space attack \cite{Doan2021_LIRA, TanS2020_BypassingBackdoor, Cheng2021_DeepFeatureSpace, li2025feat}, sample-specific attack \cite{Li2021_samplespecific}, and even attack with arbitrary target class \cite{marksman_DoanL022}, these backdoor attacks are still limited to the bonding of one trigger to one target.
The reason is that, no matter how complex the trigger generation or backdoor injection they design, the induced model behavior is predetermined and expected.
Note that, this deterministic nature is initially designed and desired.
Even for the attack with arbitrary target class \cite{marksman_DoanL022}, Doan \textit{et al} essentially inject multiple backdoors into one same model.
When the target class is chosen, the corresponding trigger is also determined through submitting class number into the conditional generation model.

Backdoor defenses \cite{0002T0SXL0M023, KolouriSPH20, ShenLTAX0M021, NC_WangYSLVZZ19, ChenCBLELMS19, ChouTP20, Tran0M18, GaoKD0ZNRK22, Tang0TZ21, JiaLCG22, 0001BSM21, KavianiSS23, ZhuPFYMAAG24}, as the opposite of attack, are inherently designed under the assumption that backdoor attacks are targeted in nature.
This fundamental assumption serves as a cornerstone for most existing defense methodologies.
Specifically, defense mechanisms are typically designed to identify and mitigate situations where compromised models exhibit consistent, predictable misclassifications towards predetermined target classes.
This underlying presumption, while valid for general backdoor attacks, will be useless involving the untargeted backdoor attack.

\subsection{Preliminaries}
In this paper, we mainly focus on backdoor attack on image classification models.
We can describe the classifier as a function $\mathcal{F}_\theta(\boldsymbol{x}): \mathcal{X} \rightarrow \mathcal{C}^k$ that learns a mapping from input images to $k$ classes.
The goal of task is to learn the parameters $\theta$ by using the training dataset $\mathcal{D}_c=\{(\boldsymbol{x}_i, y_i) \mid \boldsymbol{x}_i \in \mathcal{X}, y_i \in \mathcal{C}\}_{i=1}^{N}$, which indicates the training dataset has $N$ images.
Following the standard backdoor attack procedure, the model is trained on the combination of clean and backdoor images.
For clean images with ground-truth labels $(\boldsymbol{x}, y)$, the backdoor images $(\mathcal{T}_\xi(\boldsymbol{x}), y_t)$ are generated through a transformation function $\mathcal{T}_\xi$, where $y_t$ is usually a target label chosen by the attacker and $\mathcal{T}_\xi$ could be a trigger patching operation or a DNN-based generation model.
Through minimizing the standard cross-entropy loss function $\mathcal{L}_{CE}$, the model is trained to maximize the probability of correct predictions on the training dataset.
For clean training set, this process can be formulated as:
\begin{equation}
\theta^* = \arg\min_{\theta} \sum_{i=1}^{N} \mathcal{L}_{CE}\left(\mathcal{F}_\theta\left(\boldsymbol{x}_i\right), y_i\right)
\end{equation}
%

\noindent where $\theta$ represents the model parameters, $\theta^*$ is the optimal set of parameters.
The goal is to find the optimal parameters $\theta^*$ that minimize the cross-entropy loss over the entire training dataset, while ensuring the backdoor image is visually consistent with corresponding clean image.
Therefore, the task can be formulated as a constrained optimization problem:

\begin{equation}
\begin{aligned}
\min_{\theta,\xi} & \sum_{i=1}^{N}\upsilon \mathcal{L}_{CE}(\mathcal{F}_\theta(\boldsymbol{x}_i), y_i) + \sum_{i=1}^{M}\varsigma \mathcal{L}_{CE}(\mathcal{F}_\theta\left(\mathcal{T}_\xi(\boldsymbol{x}_i)\right), y_t) \\
\text{s.t.} & \quad \|\mathcal{T}_\xi(\boldsymbol{x}_i) - \boldsymbol{x}_i\|_\infty \leq \rho, \quad \forall i \in \{1, ..., M\}
\end{aligned}
\end{equation}

\noindent where $\upsilon$ and $\varsigma$ are used to achieve the balance between main task and backdoor task.

\begin{figure*}[htbp]
    \centering
    \includegraphics[width=0.9\linewidth]{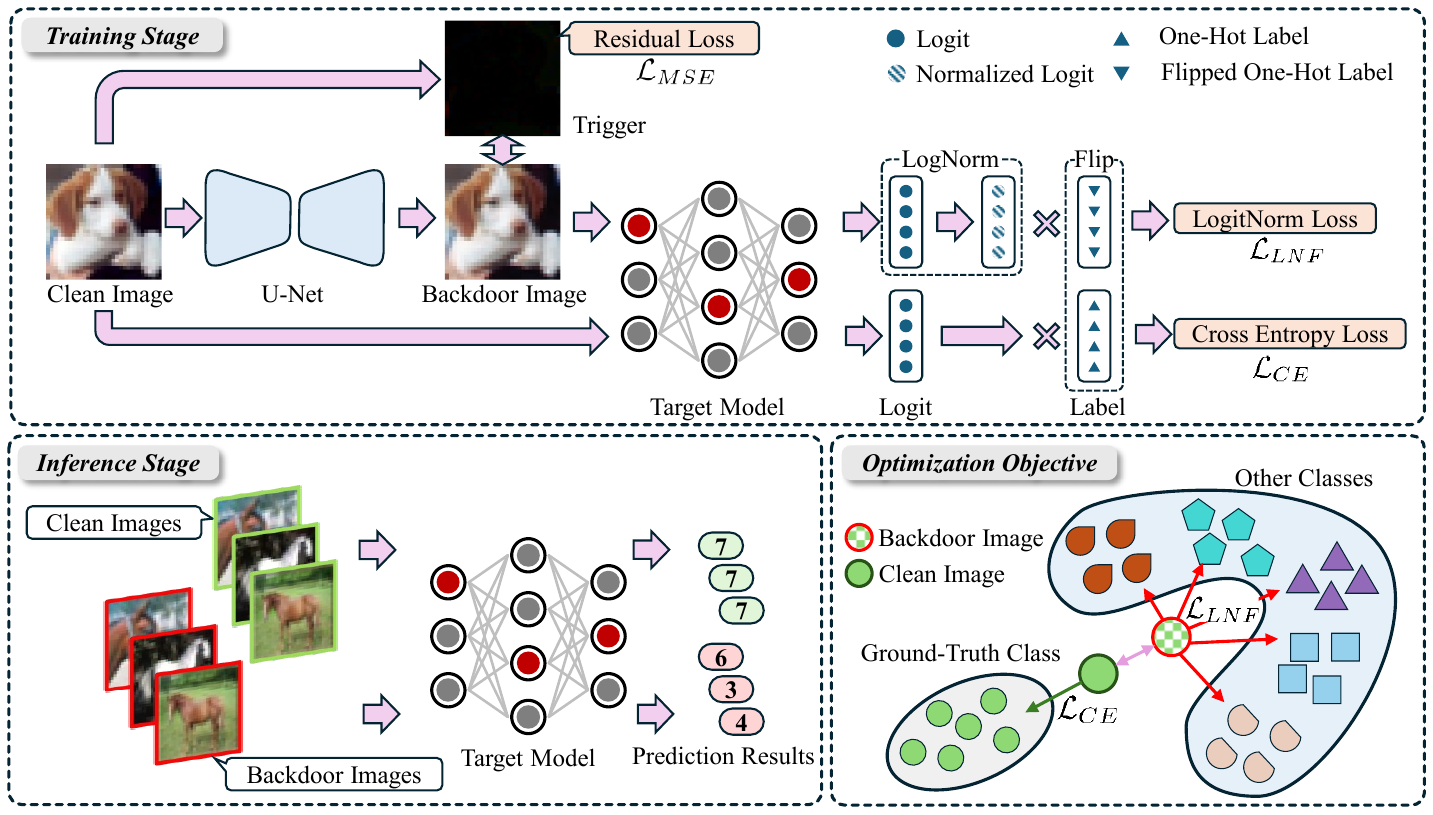}
    \caption{The overview of the proposed controlled untargeted backdoor attack.}
    \label{fig:overview}
\end{figure*}

\section{Controlled Untargeted Backdoor Attack}\label{sec:proposed_method}
\subsection{Threat Model}
\textbf{Attacker's Goal.}
The untargeted backdoor attack has two common goals and one special goal, which are stealthiness, effectiveness and dispersibility.
The first goal is to make sure the attack is imperceptible and hard to notice.
Since backdoor attack will only be triggered under specific conditions, which makes backdoor attack naturally stealthy.
For computer vision task, the stealthiness simultaneously refers to the trigger added on images should be as invisible as possible.
The second goal is to effectively perform the attack with high success rate, which means the backdoor images should be misclassified to wrong classes but clean images are classified to correct classes as normal.
The last goal is exclusive to untargeted backdoor attacks.
The dispersibility means that, to achieve the randomness of classifications, the backdoor attack is designed to predict backdoor images into different random classes as dispersive as possible.

\noindent\textbf{Attacker's Capability.}
This paper assumes the attacker has the whole control on the model training process, including images preparation, training algorithm, model architecture, which is the same threat model assumed in prior researches \cite{Nguyen2021_WanetImperceptibleWarpingBased, SahaSP2020_HiddenTriggerBackdoor, Cheng2021_DeepFeatureSpace, Doan2021_LIRA, BlindBackdoorsDeep2021Bagdasaryan, TanS2020_BypassingBackdoor}.
In this strong assumption, the attacker is the one who trained the model.
As explained in introduction, many individuals or small companies cannot train the model from scratch due to the requirement of high-quality datasets and high-performance computational devices.
They may choose to download models from third-party websites or repositories, which give the attacker opportunities to invade.
The attacker can train the backdoor model and then upload to website or other platforms for free downloading and deployment.
The compromised model is then delivered to the end users who will deploy it in applications.

\subsection{Overview}

The whole procedure of proposed CUBA is illustrated in Fig. \ref{fig:overview}, which consists of two main stages: training stage and inference stage.
During the training stage, clean images and their corresponding backdoor images (transformed by U-Net) are combined to form the backdoor training set. The optimization process incorporates three essential loss functions ($\mathcal{L}_{MSE}, \mathcal{L}_{CE}, \mathcal{L}_{LNF}$) to jointly train the target model and U-Net model.
Specifically, the Mean Squared Error loss ($\mathcal{L}_{MSE}$) is designed to control the perturbation magnitude between backdoor images and their clean counterparts, ensuring visual imperceptibility.
The standard Cross-Entropy loss ($\mathcal{L}_{CE}$) is employed to maintain normal classification performance on clean samples.
The Logit Normalization and One-Hot Flip loss ($\mathcal{L}_{LNF}$), which combines logit regularization with one-hot label flipping, serves as a crucial substitute for cross-entropy loss and constitutes the core component of our attack.
This novel loss function is specifically designed to achieve controllable untargeted attacks, enabling more flexible and stealthy backdoor interventions.
The details of logit normalization, one-hot label flipping, trigger generation and backdoor injection are elaborated in the following subsections:
\begin{itemize}
  \item Logit Normalization: To achieve untargeted attacks while maintaining control over the attack outcomes, we normalize the logits of the target model's output. This normalization ensures that the predicted probabilities for non-ground-truth classes are relatively balanced, preventing the model from consistently favoring specific incorrect classes, which may degrades to target attack. 
  \item One-Hot Label Flipping: We introduce a label flipping mechanism that transforms the original one-hot encoded labels into modified versions where the ground-truth class probability is minimized while maintaining approximately equal probabilities for all other classes. This approach enables the model to misclassify backdoored inputs without being biased toward any particular target class.
  \item Trigger Generation: The trigger patterns are generated through our U-Net architecture, which learns to produce visually imperceptible perturbations that can effectively activate the backdoor behavior. The U-Net is trained end-to-end with the target model to generate triggers that are both stealthy and efficient.
  \item Backdoor Injection: The backdoor is injected into the target model through our joint training process, where clean images and their backdoored counterparts are used simultaneously. This process ensures that the model maintains high performance on clean inputs while reliably misclassifying inputs containing the learned trigger patterns.
\end{itemize}

\subsection{Logit Normalization}
In this section, we will investigate how to mitigate the overconfidence problem, where deep neural networks trained with commonly used softmax cross-entropy loss always tends to give a high confidence on backdoor images.
We analyze the logits of model on backdoor images and clean images, and we believe the large magnitude of model can be the problem.

The output of model for an input image $\boldsymbol{x}$ is denoted as $\mathcal{F}_\theta(\boldsymbol{x}) = \boldsymbol{z}$, which is also known as the logits or output before softmax.
Making no restrictive assumptions, the logit vector $\boldsymbol{z}$ can be denoted as:

\begin{equation}
  \boldsymbol{z} = \| \boldsymbol{z} \| \cdot \tilde{\boldsymbol{z}}
\end{equation}

\noindent where $\|\boldsymbol{z}\|=\sqrt{\boldsymbol{z}_1^2+\boldsymbol{z}_2^2+\cdots+\boldsymbol{z}_k^2}$ is the \textit{Euclidean} norm (also called $\ell_2$-norm) of the logit vector $\boldsymbol{z}$, and $\tilde{\boldsymbol{z}}$ is the unit vector with the same direction of $\boldsymbol{z}$.
Therefore, we can take the $\|\boldsymbol{z}\|$ and $\boldsymbol{z}$ as  the \textit{magnitude} and \textit{direction} of the logit vector $\boldsymbol{z}$.

During inference stage, if we only care the predicted class label without probability values, we can use $\operatorname*{arg\,max}$ directly on logits $\boldsymbol{z}=\{z_i \mid i=1,2,3, \dots, k\}$ to get the predicted label $c=\operatorname*{arg\,max}_i (z_i)$.
If $z_i$ multiplies with a constant value $\lambda$ ($
\lambda >0 $), the resulting classification remains invariant.
This can be formally expressed as:
\begin{equation}\label{eq:argmax1}
  \operatorname*{arg\,max}_i (z_i) = \operatorname*{arg\,max}_i (\lambda z_i).
\end{equation}

In practice, logit vectors typically undergo a softmax transformation as

\begin{equation}\label{eq:softmax}
  \text{softmax}(z_i) = \frac{e^{z_i}}{\sum_{j=1}^k e^{z_j}}
\end{equation}

\noindent before the argmax operation.
Take softmax into consideration, given any scalar $\lambda > 1$, if $c=\operatorname*{arg\,max}_i(z_i)$, then $  \text{softmax}(\lambda \cdot z_i) > \text{softmax}(z_i)$ always holds.
Once we scale the logit vector $\boldsymbol{z}$ by scalar $\lambda$, all components change proportionally.
Therefore, scaling the magnitude $\|\boldsymbol{z}\|$ will not affect the direction of $\boldsymbol{z}$:
\begin{equation}\label{eq:argmax2}
    \operatorname*{arg\,max}_i (\text{softmax}(\lambda \boldsymbol{z})_i) = \operatorname*{arg\,max}_i (\text{softmax}(\boldsymbol{z})_i)
\end{equation}
\noindent which also means the final prediction result remains unchanged.
Furthermore, we analyze the final influence on optimization objective function.
In classification task, cross entropy is widely used to calculate the distance between predicted labels and ground-truth classes:
\begin{equation}\label{eq:ce_std}
\mathcal{L}_{CE}(\boldsymbol{z},y) = - \sum_{t=1}^{k}\mathbb{I}\{y=t\} \cdot \log \frac{e^{z_t}}{\sum_{j=1}^k e^{z_j}}. 
\end{equation}

During backdoor training process, minimizing cross entropy loss on pairs of backdoor images and target classes will constantly increase the magnitude $z_t$, which finally encourage the backdoor model to an overconfident prediction.
This characteristic is the key of backdoor attack because it establishes a strong connection between backdoor image and target label, pushing the attack success rate to a high level.
However, this characteristic also leaves footprints in pixel space or even latent space, which gives the defense methods opportunities to detect and mitigate backdoor attacks.



To alleviate this problem, the straight forward solution is to control the magnitude of logit vector under a certain threshold $\zeta$, which can be achieved by imposing an $\ell_2$-norm constraint on the logit outputs.
This is conducive to prevent the model from producing over-confident predictions and reduces the risk of numerical instability.
The final cross entropy based optimization objective function can be formulated as:

\begin{equation}
\begin{aligned}
  \min & \mathop{\mathbb{E}}\limits_{(\boldsymbol{x},y)\in \mathcal{D}}\left[\mathcal{L}_{CE} \left(\boldsymbol{z}, y \right) \right] \\
  \text{s.t.} & \quad  \|\boldsymbol{z}\|_2=\|\boldsymbol{z\|} \leq \zeta
\end{aligned}
\end{equation}

This normalization applied on logit is called \textbf{Logit Normalization (LogNorm)}, which keeps the direction of logit vector without constantly increasing the magnitude of logit vector.
Instead of applying softmax cross-entropy loss directly to the original logit outputs, we first normalize the logit vector.
The objective function with LogitNorm can be mathematically formulated as:


\begin{equation}\label{eq:lognorm}
  \mathcal{L}_{LN}(\boldsymbol{z}, y) = - \sum_{t=1}^{k}\mathbb{I}\{y=t\} \cdot \log \frac{e^{\tilde{\boldsymbol{z}}_t/\tau}}{\sum_{j=1}^k e^{\tilde{\boldsymbol{z}}_j/\tau}}
\end{equation}
\noindent where $\tilde{\boldsymbol{z}}=\boldsymbol{z}/(\|\boldsymbol{z}\|+\epsilon)$ is the normalized logit vector.
$\epsilon$ is a small constant for numerical stability, and $\tau$ denotes the temperature parameter controlling the softness of probability distribution.
$\mathbb{I}\{y=t\}$ is the indicator function that equals $1$ when the label is the ground truth label and $0$ otherwise.

\begin{figure}[!t]
    \centering
    \includegraphics[width=\linewidth]{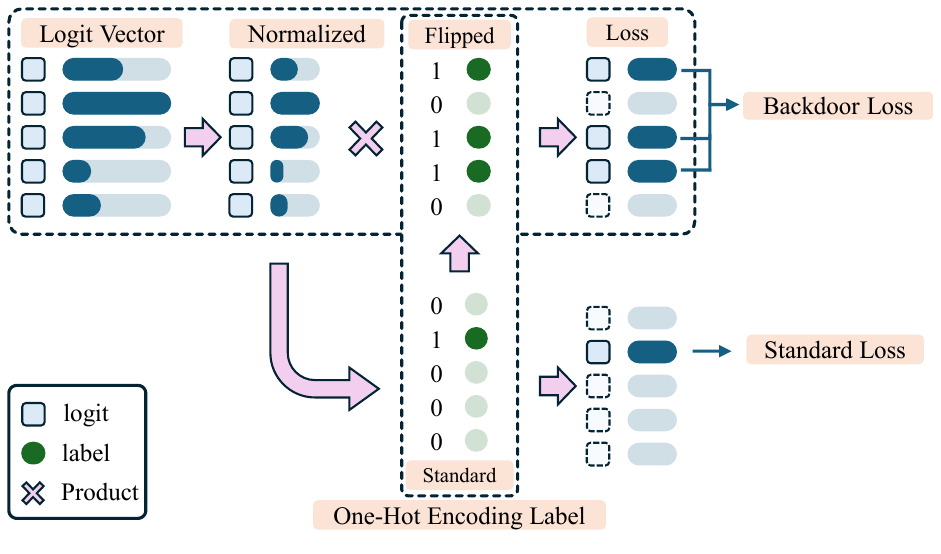}
    \caption{The illustration of Logit Normalization and Flipped One-Hot Enconding.}
    \label{fig:lognorm_flip}
\end{figure}

\subsection{Flipped One-Hot Encoding}
Only suppressing the magnitude is not enough to manipulate the model's prediction results.
Therefore, we propose a \textbf{Flipped One-Hot Encoding (FOHE)} method to implement controlled untargeted backdoor attacks.
The FOHE mechanism performs a direct inversion of the binary values in the encoding, transforming the original label representation.
Building upon this transformation, we formulate the logit normalization loss as follows:

\begin{equation}\label{eq:lognorm_flip_FRA}
    \mathcal{L}_{LNF}(\boldsymbol{z}, y) = - \sum_{t=1}^{k}\left[1 - \mathbb{I}\{y=t\}\right] \cdot \log \frac{e^{\tilde{\boldsymbol{z}}_t/\tau}}{\sum_{j=1}^k e^{\tilde{\boldsymbol{z}}_j/\tau}}
\end{equation}

\noindent where $y$ is the ground-truth label, and $1- \mathbb{I}\{y=t\}$ serves as a standard one-hot encoding function with flipped labels.

For the proposed two attack variants of CUBA, Eq. \ref{eq:lognorm_flip_FRA} can represent the FRA, but for NRA, the random target classes are constrained in a subset of classes $\mathcal{S}$ selected by the attacker.
For NRA, the logit normalization loss is as follows:

\begin{equation}\label{eq:lognorm_flip_NRA}
    \mathcal{L}_{LNF}(\boldsymbol{z}, y, \mathcal{S}) = - \sum_{t=1}^{k}\mathbb{I}\{y \in \mathcal{S}\}\cdot \log \frac{e^{\tilde{\boldsymbol{z}}_t/\tau}}{\sum_{j=1}^k e^{\tilde{\boldsymbol{z}}_j/\tau}}
\end{equation}

\noindent where obviously the attacker-chosen target classes do not contain the ground-truth class $y \notin \mathcal{C}\cap\mathcal{S}$.

%
%
%

\subsection{Trigger Generation and Backdoor Injection}
Following those backdoor attacks \cite{BackdoorAttackImperceptible2021Doan, Doan2021_LIRA, Li2021_samplespecific} that using a image processing model as the trigger generator, we utilize a U-Net model \cite{RonnebergerFB2015_UNet} to generate dynamic trigger for backdoor images.
In this paper, the U-Net model is trained simultaneously with the target model during training stage.
To make the trigger as imperceptible as possible, we calculate the Mean Squared Error (MSE) of clean image and corresponding backdoor image as the MSE loss $\mathcal{L}_{MSE}$.

As shown in Algorithm \ref{alg:main}, except for standard cross-entropy loss for main task, we apply logit normalization with flipped one-hot encoding for backdoor task.
During each iteration, all clean images $\boldsymbol{x}$ are transformed by U-Net model as the backdoor images $\mathcal{T}_{\xi}(\boldsymbol{x})$ and concatenated with the clean images.
The clean images and backdoor images are submitted to target model for predictions.
Based on the logit vectors of the target model, we split the logit vectors into clean logit and backdoor logit.
For clean images, the loss function is still standard cross entropy loss with one-hot encoding label.
For backdoor images, the loss function changes to logit normalization loss with flipped one-hot encoding label.
The final loss function is formulated as:
\begin{equation}
  \mathcal{L}_{total} = \boldsymbol{\alpha} \cdot \mathcal{L}_{MSE} + \boldsymbol{\beta} \cdot \mathcal{L}_{CE} + \boldsymbol{\gamma} \cdot \mathcal{L}_{LNF}
\end{equation}
\noindent where $\boldsymbol{\alpha}$, $\boldsymbol{\beta}$ and $\boldsymbol{\gamma}$ is three hyper-parameters used to balance the model utility, attack performance and trigger stealthiness.

\begin{algorithm}[t]
\caption{The training process of CUBA}
\label{alg:main}
\begin{algorithmic} 
  \Require Initialized parameters $\theta$ and $\xi$; The clean training set $\mathcal{D}_c$; The training iterations $I$; The batch size $B$; The learning rate $lr$; The attacker chosen labels $\mathcal{S}$
  \Ensure Optimized parameters $\theta$ and $\xi$
  \While {$i \in I$}
  \State $i \leftarrow i+1$; mini-batch sampling $\{(\boldsymbol{x}_j, y_j)\}_{j=1}^{B} \subset \mathcal{D}_c$
  \For {each image $(\boldsymbol{x}, y)$ in mini-batch }
      \State $\boldsymbol{x}^* = \mathcal{T}_\xi(\boldsymbol{x})$
      \Comment {Generate backdoor image}
      \State $\mathcal{L}_{MSE} = MSE(\boldsymbol{x}^*, \boldsymbol{x})$
      \Comment {Calculate MSE Loss}
      \State $\boldsymbol{z} = \mathcal{F}_\theta(\boldsymbol{x}), \boldsymbol{z}^* = \mathcal{F}_\theta(\boldsymbol{x}^*)$
      \Comment {Get logit vector}
      \State $\mathcal{L}_{CE}(\boldsymbol{z}, y)$ 
      \Comment {Calculate logit standard (Eq. \ref{eq:ce_std}) }
      \State $\mathcal{L}_{LNF}(\boldsymbol{z}^*, y, \mathcal{S})$
      \Comment {Calculate logit norm (Eq. \ref{eq:lognorm_flip_FRA}) }
      \State $\mathcal{L}_{total} \leftarrow \boldsymbol{\alpha} \cdot \mathcal{L}_{MSE} + \boldsymbol{\beta} \cdot \mathcal{L}_{CE} + \boldsymbol{\gamma} \cdot \mathcal{L}_{LNF}$
      \State $\theta \leftarrow \theta - lr \cdot \bigtriangledown_\theta \cdot \mathcal{L}_{total}$
      \State $\xi \leftarrow \xi - lr \cdot \bigtriangledown_\xi \cdot \mathcal{L}_{total}$
  \EndFor
  \EndWhile
\end{algorithmic}
\end{algorithm}


\section{Experiments}\label{sec:experiments}
\begin{table*}[htbp]
\renewcommand{\arraystretch}{1} 
\centering
\caption{Experimental results of the proposed CUBA with FRA and NRA variants. $\uparrow$ indicates the higher the better.}
\begin{tabularx}{\textwidth}{>{\centering\arraybackslash}p{3cm}c>{\centering\arraybackslash}X>{\centering\arraybackslash}X>{\centering\arraybackslash}X>{\centering\arraybackslash}X>{\centering\arraybackslash}X>{\centering\arraybackslash}X}
\toprule
Model                           & Dataset                   & Attack & ASR(\%)$\uparrow$    & DS(\%)$\uparrow$  & CA(\%)$\uparrow$        & Baseline                \\ \cmidrule{1-7}
\multirow{3}[3]{*}{SimpleCNN}      & \multirow{3}[3]{*}{MNIST}    & FRA    & 99.99$\pm$0.00 & 90.74$\pm$0.15 & 99.45$\pm$0.07 & \multirow{3}[3]{*}{99.58\%} \\ \cmidrule(lr){3-3} \cmidrule(lr){4-6}
								&                           & NRA-5  & 99.98$\pm$0.01 & 91.59$\pm$0.12 & 99.54$\pm$0.05 &                         \\ \cmidrule(lr){3-3} \cmidrule(lr){4-6}
								&                           & NRA-2  & 99.96$\pm$0.02 & 93.43$\pm$0.09 & 99.53$\pm$0.04 &                         \\ \cmidrule{1-7}
\multirow{3}[3]{*}{PreActResNet18} & \multirow{3}[3]{*}{GTSRB}    & FRA    & 99.94$\pm$0.01 & 98.52$\pm$0.05 & 98.36$\pm$0.02 & \multirow{3}[3]{*}{98.69\%} \\ \cmidrule(lr){3-3} \cmidrule(lr){4-6}
								&                           & NRA-5  & 99.86$\pm$0.02 & 98.97$\pm$0.01 & 98.41$\pm$0.01 &                         \\ \cmidrule(lr){3-3} \cmidrule(lr){4-6}
								&                           & NRA-2  & 99.89$\pm$0.02 & 99.13$\pm$0.02 & 98.43$\pm$0.01 &                         \\ \cmidrule{1-7}
\multirow{3}[3]{*}{ResNet18}       & \multirow{3}[3]{*}{CIFAR10}  & FRA    & 99.79$\pm$0.02 & 92.70$\pm$0.08 & 93.49$\pm$0.31 & \multirow{3}[3]{*}{93.91\%} \\ \cmidrule(lr){3-3} \cmidrule(lr){4-6}
								&                           & NRA-5  & 99.77$\pm$0.10 & 95.81$\pm$0.14 & 93.66$\pm$0.27 &                         \\ \cmidrule(lr){3-3} \cmidrule(lr){4-6}
								&                           & NRA-2  & 99.71$\pm$0.08 & 98.39$\pm$0.06 & 93.87$\pm$0.23 &                         \\ \cmidrule{1-7}
\multirow{3}[3]{*}{ResNet18}       & \multirow{3}[3]{*}{CIFAR100} & FRA    & 98.84$\pm$0.05 & 97.06$\pm$0.03 & 74.60$\pm$0.22 & \multirow{3}[3]{*}{74.98\%} \\ \cmidrule(lr){3-3} \cmidrule(lr){4-6}
								&                           & NRA-5  & 98.71$\pm$0.03 & 97.59$\pm$0.05 & 74.67$\pm$0.10 &                         \\ \cmidrule(lr){3-3} \cmidrule(lr){4-6}
								&                           & NRA-2  & 97.68$\pm$0.02 & 97.61$\pm$0.02 & 74.64$\pm$0.18 &                         \\ \bottomrule
\end{tabularx}
\label{table:exp_res}
\end{table*}

\subsection{Datasets and Models}
To comprehensively evaluate the effectiveness of our proposed CUBA, we conduct extensive experiments on three widely-used datasets with their corresponding model architectures:
For MNIST \cite{deng2012mnist}, we employ a SimpleCNN architecture, which consists of two convolutional layers followed by three fully connected layers.
For GTSRB \cite{GTSRB_Houben2013}, we utilize PreActResNet18 \cite{HeZRS16_PreActResNet18}, a variant of ResNet that places batch normalization and ReLU activation before convolution operations.
For CIFAR10 \cite{krizhevsky2009learning_cifar10}, we implement the standard ResNet18 \cite{HeZRS16} architecture for this dataset due to its proven effectiveness in handling complex image classification tasks.

\subsection{Evaluation Metrics}
To systematically evaluate the performance of CUBA, we employ two primary metrics: Attack Success Rate (ASR) and Dispersibility Score (DS).

\noindent \textbf{Attack Success Rate.} The ASR is calculated as the ratio between successfully attacked samples and the total number of submitted samples:
\begin{equation}
\text{ASR} = \frac{\sum_{i=1}^{N} \mathbb{I}(\mathcal{F}_\theta(\boldsymbol{x}_i^*) \neq y_i)}{N}.
\end{equation}
For Full-Range Attack (FRA), a successful attack occurs when the prediction differs from the true label. In Narrow-Range Attack (NRA), success is achieved when the prediction falls within a dynamic target range $\mathcal{R}_i$ defined for each sample $i$.

\begin{figure}[htbp]
    \centering
    \includegraphics[width=0.9\linewidth]{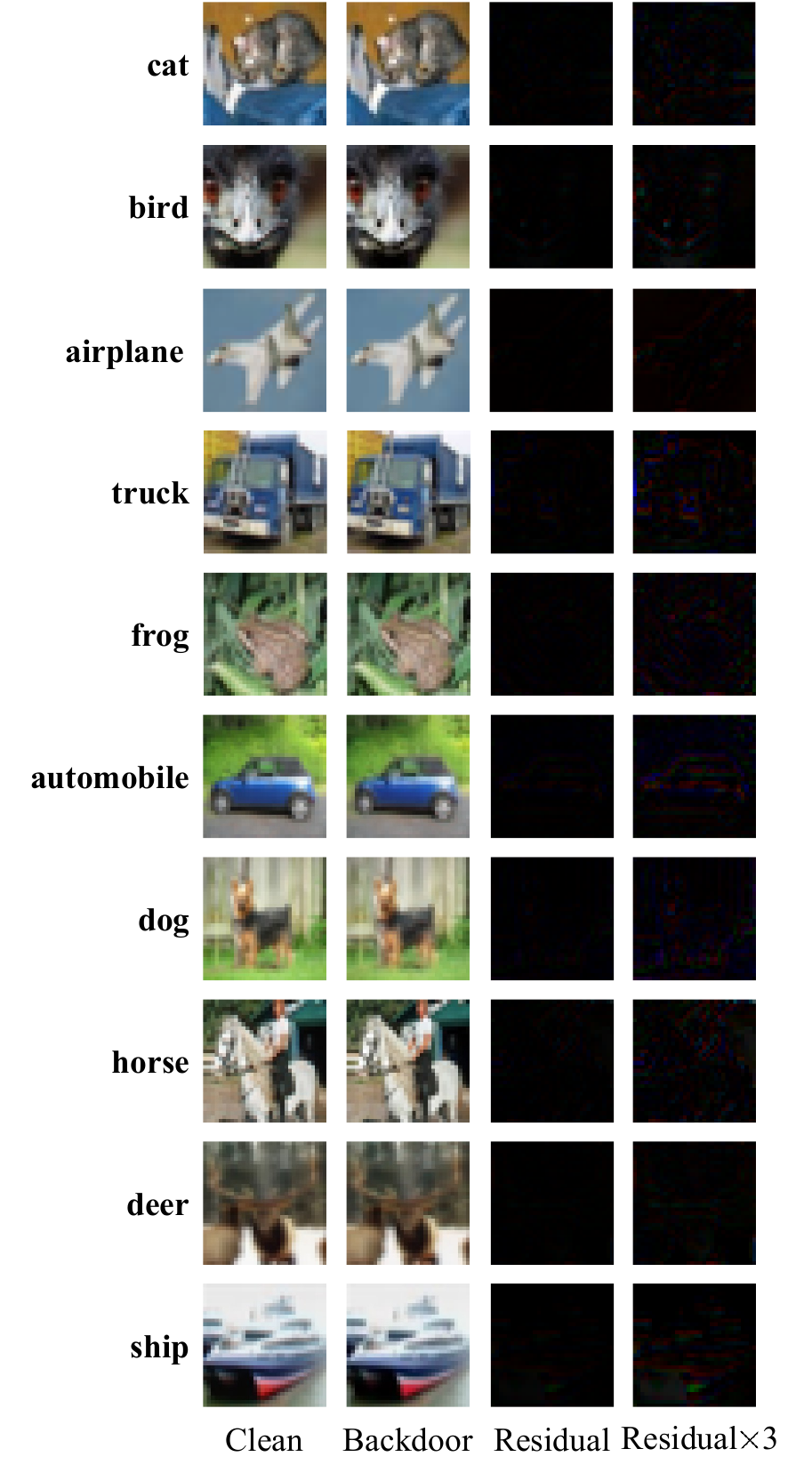}
    \caption{Samples of clean images, backdoor images and residuals in CIFAR10 dataset.}
    \label{fig:show}
\end{figure}

\noindent \textbf{Dispersibility Score.}
To evaluate the uniformity of attack outcomes across target classes, we introduce the Dispersibility Score (DS). This metric quantifies how evenly the attacked samples are distributed among the designated target classes:
\begin{equation}
\text{DS} = 1 - \sqrt{\frac{\sum_{j \in \mathcal{H}} (p_j - \frac{1}{|\mathcal{H}|})^2}{|\mathcal{H}|}}
\end{equation}
where $\mathcal{H}$ represents the set of target classes, $p_j$ is the proportion of successful attacks that resulted in class $j$, and $|\mathcal{H}|$ is the cardinality of the target set.
A higher dispersibility score (closer to 1) indicates more uniform distribution across target classes, while a lower score suggests bias toward specific classes.

\subsection{Attack Performance}
\begin{figure}[htbp]
    \centering
    \includegraphics[width=\linewidth]{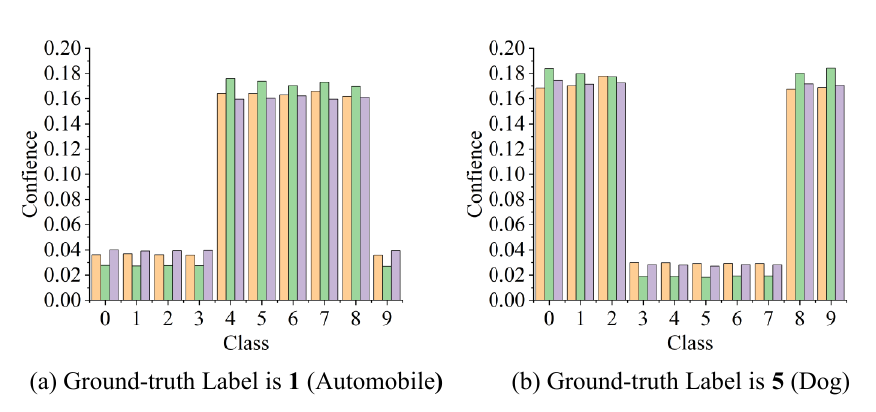}
    \caption{The confidence distribution of the compromised model on backdoor images. We show the value from 3 images for each ground-truth label.}
    \label{fig:confidence}
\end{figure}
The attack performance is evaluated with Attack Success Rate (ASR), Dispersibility Score (DS) and Clean Accuracy (CA).
Note that, for NRA, we set the number of target classes to 5 and 2, and for FRA, the number of target classes is equal to the number of all output classes.
We also report the classification performance of the clean model without attack as the baseline for comparison.
The results are shown in Table \ref{table:exp_res} and we show the clean images, backdoor images and residuals between them in Fig. \ref{fig:show}.
All methods achieved exceptional performance with ASR exceeding 99.90\% (MNIST), while ensures dispersible predictions (90.74\%-99.13\%) and clean accuracy (as high as baseline).
The performance remains robust on more complex datasets like GTSRB, where PreActResNet18 exhibited strong results with ASR higher than 99.8\% and DS higher than 98.9\%.
The FRA variant showed marginally better ASR across most scenarios, while NRA-2 demonstrated superior utility maintenance capabilities.
It demonstrate the effectiveness of the proposed CUBA approach with both FRA and NRA variants across different architectures and datasets while ensuring the model normal utility.
To evaluate the performance on larger dataset with more classes, we also report the results on CIFAR100.
The slight performance decline observed on CIFAR100 (around 74\%) is caused by the limitations of ResNet18 when scaling to larger datasets.
These experiments are conducted on the platform with single GPU NVIDIA RTX 3090, Ubuntu 22.04 LTS.
The hyper-parameters $\boldsymbol{\alpha}$, $\boldsymbol{\beta}$ and $\boldsymbol{\gamma}$ is set to 1, 1, 5, respectively.

\subsection{Attack Robustness}
\begin{figure}[htbp]
    \centering
    \begin{subfigure}[b]{0.45\linewidth}
        \centering
        \includegraphics[width=\linewidth]{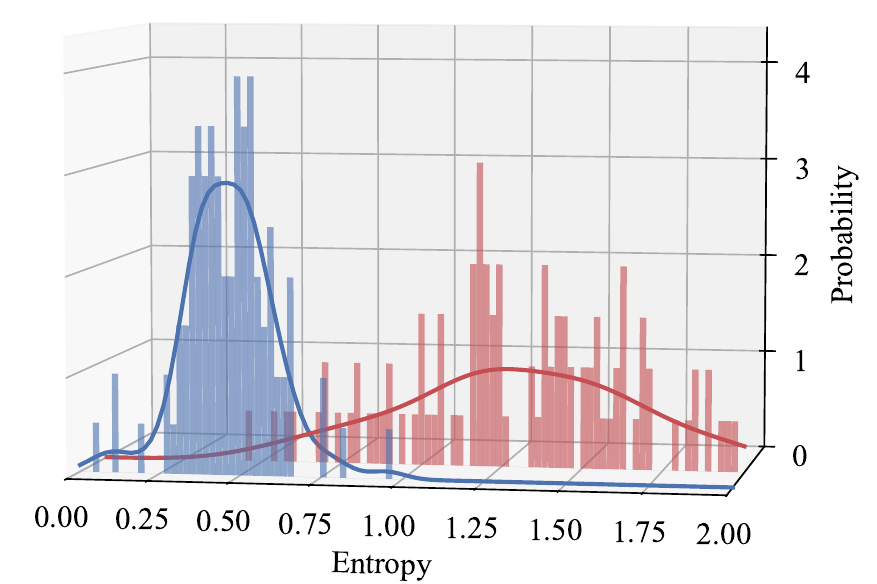}
        \caption{CIFAR10}
        \label{fig:strip1}
    \end{subfigure}
    \hfill
    \begin{subfigure}[b]{0.43\linewidth}
        \centering
        \includegraphics[width=\linewidth]{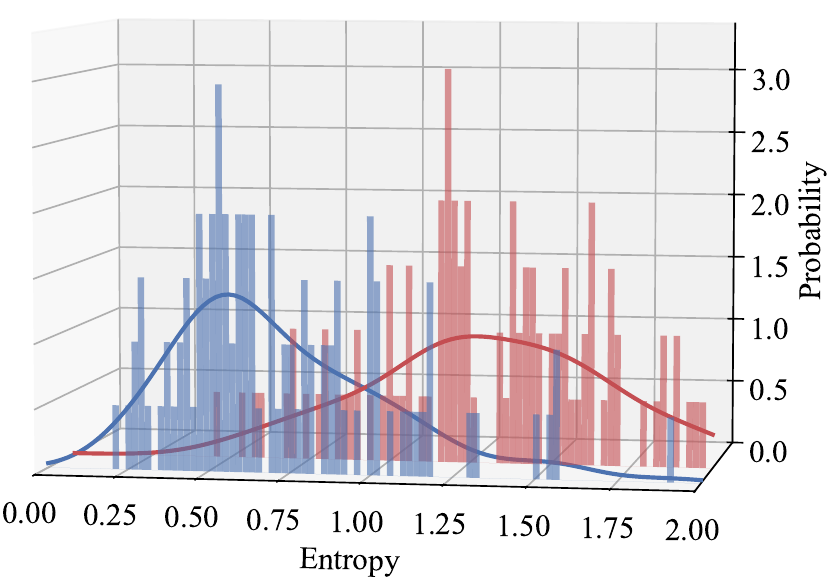}
        \caption{GTSRB}
        \label{fig:strip2}
    \end{subfigure}
    \caption{The distributions of entropy calculated by STRIP on CIFAR10 and GTSRB.
    The red and blue histogram indicate backdoor and clean images respectively. }
    \label{fig:strip}
\end{figure}

\textbf{Resistance to STRIP.}
STRIP \cite{GaoKD0ZNRK22} is a plug and play backdoor detection method which assumes that backdoor model produces consistent predictions for backdoor images, exhibiting resilience to perturbations.
This core of this detection mechanism is to analyze the classification entropy after superimposing random patterns onto the input images.
The entropy probability distributions of clean and backdoor images is illustrated in Fig. \ref{fig:strip}.
It shows that, the entropy of backdoor images are overlapped or even higher than that of clean images.
Therefore, STRIP cannot distinguish the backdoor merely relying on the entropy.
We believe the untargeted attack characteristic causes the high entropy of backdoor images, which breaks the targeted attack assumption that STRIP relies on.

\begin{figure}[htbp]
    \centering
    \includegraphics[width=\linewidth]{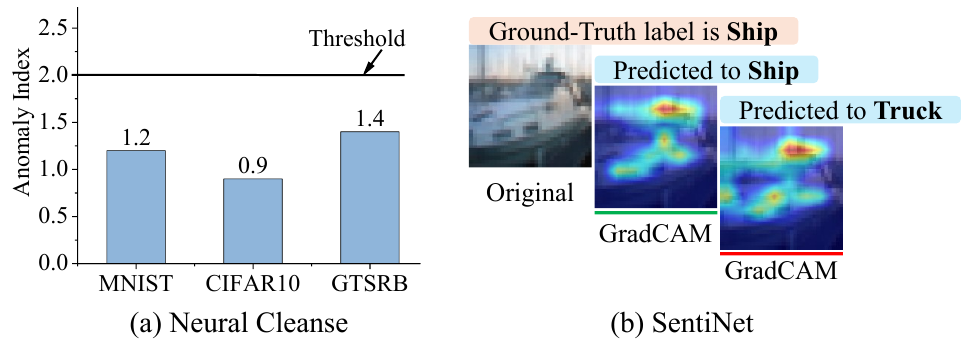}
    \caption{The detection results of Neural Cleanse and SentiNet.}
    \label{fig:nc_cam}
\end{figure}

\noindent \textbf{Resistance to Neural Cleanse.}
Neural Cleanse (NC) \cite{NC_WangYSLVZZ19} is a classic reverse engineer based backdoor detection method.
It employs trigger reconstruction through reverse engineering to identify potential backdoors within the model architecture.
We present our experimental results in Fig. \ref{fig:nc_cam} (a), demonstrating the resistance across three benchmark datasets.
The detection efficacy is primarily governed by the anomaly index, with a predefined threshold value of 2.0 following settings in \cite{NC_WangYSLVZZ19}.
The anomaly indices across all three datasets fell below the predefined threshold, which can be attributed to the inherent limitations of Neural Cleanse when facing complex trigger patterns.
Specifically, Neural Cleanse exhibits reduced effectiveness in scenarios involving large-scale or discrete trigger patterns.
The U-Net-generated triggers prove particularly challenging to reverse engineer, consequently impacting the reliability of the detection results.

\noindent \textbf{Resistance to SentiNet.}
The SentiNet defense mechanism \cite{ChouTP20} leverages input saliency maps to identify potential trigger regions within input samples.
This method first utilizes GradCAM \cite{selvaraju2017grad} to locate suspicious regions which may contain trigger.
Then these suspicious regions are cropped and pasted onto other images to obtain the corresponding prediction results.
If most predictions point to the same labels, the regions are considered to contain trigger. 
As illustrated in Fig. \ref{fig:nc_cam} (b), the saliency maps of clean image and backdoor image is very similar and hard to distinguish.
For general backdoor attacks, localized patterns are usually detectable by SentiNet.
However, the proposed CUBA generates semantically consistent triggers that do not associate to any specific target class.
Moreover, the trigger is strategically distributed throughout the whole clean image through the optimized U-Net model, which makes the conventional saliency map based methods hard to detect.

\begin{table}[htbp]
  \setlength\tabcolsep{4pt}
  \renewcommand{\arraystretch}{0.9} 
  \centering
  \caption{The ASR and CA of the backdoor model after pruning and distillation.}
    \begin{tabular}{cccccc}
    \toprule 
    \multicolumn{6}{c}{Fine Pruning} \\
    \toprule
     Rate & 10\%  & 30\%  & 50\%  & 70\%  & 90\% \\
\cmidrule{1-6}           ASR   & 99.61\% & 88.44\% & 64.05\% & 48.98\% & 17.48\% \\
\cmidrule{1-6}           CA    & 92.41\% & 86.59\% & 76.68\% & 34.78\% & 25.28\% \\
    \toprule
    \multicolumn{6}{c}{Neural Attention Distillation} \\
    \toprule
 Epoch & 2     & 4     & 6     & 8     & 10 \\
\cmidrule{1-6}          ASR   & 81.39\% & 80.91\% & 79.22\% & 76.40\% & 66.89\% \\
\cmidrule{1-6}          CA    & 93.61\% & 92.07\% & 89.85\% & 87.99\% & 77.34\% \\
    \bottomrule
    \end{tabular}%
  \label{tab:fp_nad}%
\end{table}%

\noindent \textbf{Resistance to Fine-pruning}
The key of Fine-pruning \cite{0017DG18} is to prune the reluctant neurons that are not activated when inputting clean images.
Fine-pruning believe these dormant neurons are backdoor neurons that needs to be removed.
We evaluate the robustness of our attack against neuron pruning using ResNet18 on CIFAR10, specifically focusing on the pruning operations in the final convolutional layer.
As illustrated in Table \ref{tab:fp_nad}, our method demonstrates remarkable resilience against this defense mechanism.
The consistent synchronization between the ASR and the CA indicates that backdoor neurons are deeply entangled with clean neurons.
This tight bounding makes it difficult to remove the backdoor without affecting normal classification performance. 

\noindent \textbf{Resistance to Neural Attention Distillation}
Like normal distillation methods, Neural Attention Distillation (NAD) \cite{LiLKLLM21} leverages a teacher model to guide the fine-tuning of the backdoor student model through a small clean dataset.
As shown in Table \ref{tab:fp_nad}, NAD not only reduces the Attack Success Rate (ASR) but also significantly lowers the Clean Accuracy (CA).
This suggests that while NAD is effective in diminishing the impact of backdoor attack, it does so at the cost of reducing the performance of model on clean images.
It demonstrates that, CUBA does not affect the model utility, in fact, heavily relies on the model's clean classification capability.
In other words, the higher the CA, the higher the ASR, and vice versa.

\subsection{Compared with Closely Related Work}
\begin{figure}[htbp]
    \centering
    \includegraphics[width=0.85\linewidth]{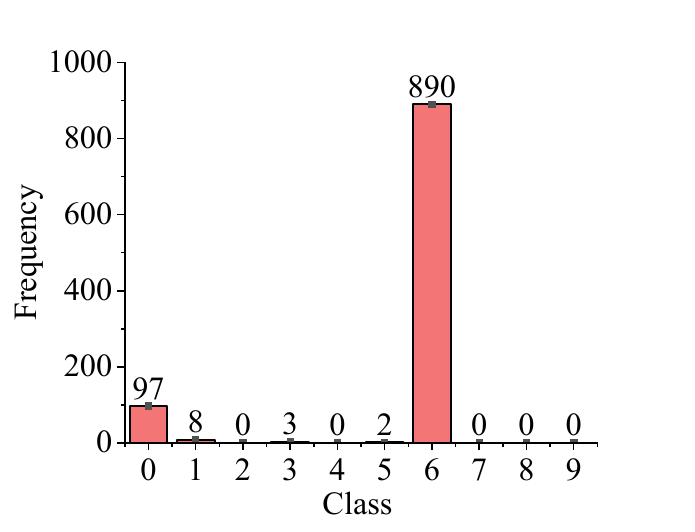}
    \caption{The distributions of predictions results of backdoored model from \cite{XueWNZZL24}. The ground-truth class of these 1,000 images is 1.}
    \label{fig:old_uba}
\end{figure}

To the best of our knowledge, the only closely related work in this domain is the untargeted backdoor attack proposed \cite{XueWNZZL24}.
However, this approach presents a significant limitation: by solely modifying the loss function to facilitate the backdoor attack, it inadvertently causes the untargeted attack to degrade into a targeted one.
We report the confidence distribution of work \cite{XueWNZZL24} in Fig. \ref{fig:old_uba}.
Among 1,000 test images, 890 backdoor images from class 1 are misclassified to class 6 under the simple untargeted backdoor attack \cite{XueWNZZL24}.
Meanwhile, the dispersibility score is 0.7325, which is very close to 0.7 (the minimum value of DS on 10 classes).
The following derivation demonstrates that a dispersibility score of 0.7 corresponds to predictions being almost entirely concentrated in a single class.
For a $H$-classes model, 
\begin{equation}
  p_j = \frac{1}{H} \quad \text{for all } j \in \{1, 2, \ldots, H\}
\end{equation}

if we assume all predictions fall into a single class, says class $k$: 

\begin{equation}
p_k = 1 \quad \text{and} \quad p_j = 0 \quad \text{for } j \neq k.
\end{equation}

The variance is calculated by
\begin{equation}
\text{variance} = \frac{1}{H} \sum_{j=1}^{H} \left( p_j - \frac{1}{H} \right)^2
\end{equation}

\noindent substituting the values:

\begin{equation}
  \text{variance} = \frac{1}{H} \left[ \left(1 - \frac{1}{H}\right)^2 + \sum_{j \neq k} \left(0 - \frac{1}{H}\right)^2 \right]
\end{equation}

\noindent continued by simplifying:

\begin{equation}
  \text{variance} = \frac{(H - 1)(H - 1 + 1)}{H^3} = \frac{H - 1}{H^2}
\end{equation}

Therefore, the final dispersibility score is calculated as ($H=10$):

\begin{equation}
  DS = 1 - \sqrt{\frac{H - 1}{H^2}} = 1 - \frac{\sqrt{9}}{10} = 0.7
\end{equation}

It shown that, their method \cite{XueWNZZL24} yields an extremely low dispersibility score, nearly aligning with that of a targeted attack.
We attribute this phenomenon to an inherent model bias, that neural networks naturally tend to seek a ``shortcut" solution \cite{GeirhosJMZBBW20} during optimization.
In this case, by altering the loss function to push the backdoor images away from its original class, the model consistently redirects backdoor inputs to the nearest alternative class rather than towards true untargeted misclassification across multiple classes.
Through this experimental results, we identified a critical limitation in their proposed approach \cite{XueWNZZL24} of using loss function modification for untargeted backdoor attack.
Specifically, solely modifying the cross-entropy loss without considering overconfidence problem cannot perform a successful untargeted backdoor attack.
This behavior emerges because the continuous optimization of the modified cross-entropy loss function \cite{XueWNZZL24} inadvertently creates a pattern where the model satisfies the optimization objective by simply selecting the closest alternative class for misclassification.
Eventually, despite being designed for untargeted attacks, their approach \cite{XueWNZZL24} effectively collapses into a targeted attack scenario where backdoor samples are predominantly misclassified into a single target class.

\section{Conclusion}\label{sec:conclusion}
This paper proposes CUBA, a novel backdoor attack paradigm that enables controlled untargeted misclassifications.
CUBA fundamentally differs from traditional targeted backdoor attacks by allowing random misclassifications within predefined class ranges through full range attack and narrow range attack variants. 
Through the proposed logit normalization with one-hot flipping, CUBA successfully achieves uniform distribution of misclassifications while maintaining performance on clean data. 
This proposed attack reveals new vulnerabilities in model security, and we hope it can promote the defenses for untargeted backdoor attacks.

\bibliographystyle{IEEEtran}
\bibliography{ref}

\begin{IEEEbiography}[{\includegraphics[width=1in,height=1.25in,clip,keepaspectratio]{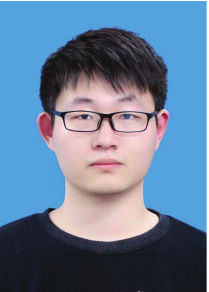}}]{Yinghao Wu} received the M.S. degree in computer science and technology from Nanjing University of Aeronautics and Astronautics in 2023. He is currently pursuing the Ph.D. degree with the College of Computer Science and Technology, Nanjing University of Aeronautics and Astronautics, Nanjing, China. His research interests include artificial intelligence security and privacy.
\end{IEEEbiography}

\vskip -2.75\baselineskip plus -1fil
\begin{IEEEbiography}[{\includegraphics[width=1in,height=1.25in,clip,keepaspectratio]{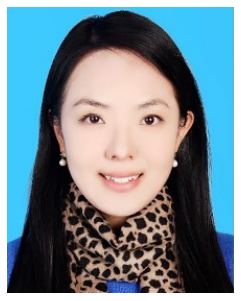}}]{Liyan Zhang} is currently a Professor with the School of Computer Science and Technology, Nanjing University of Aeronautics and Astronautics, Nanjing, China. She received the Ph.D. degree in computer science from the University of California, Irvine, Irvine, CA, USA, in 2014. Her research interests include multimedia analysis, computer vision, and deep learning. Dr. Zhang received the Best Paper Award from the International Conference on Multimedia Retrieval (ICMR) 2013, the Best Student Paper Award from the International Conference on Multimedia Modeling (MMM) 2016, the Best Student Paper Award from International Conference on Internet Multimedia Computing and Service (ICIMCS) 2017, and the Best Paper Award from ACM Multimedia Asia conference (MM Asia) 2021. 
\end{IEEEbiography}

\end{document}